\documentclass[article,aps,twocolumn,showpacs,amsmath,amssymb,superscriptaddress,nofootinbib,floatfix]{revtex4-2}

\usepackage{graphicx}
\usepackage{dcolumn}
\usepackage{bm}
\usepackage[colorlinks=true,linkcolor=blue,citecolor=blue,urlcolor=blue]{hyperref}
\usepackage[normalem]{ulem}

\usepackage{array,booktabs,tabularx}
\usepackage[caption=false]{subfig}
\usepackage[USenglish]{babel}
\usepackage{braket}
\usepackage{xcolor}
\usepackage{amsfonts}
\usepackage{amssymb}
\usepackage[normalem]{ulem}
\usepackage{amsmath}
\usepackage[utf8]{inputenc}
\usepackage{bibunits}
\defaultbibliographystyle{apsrev4-2}

\makeatletter

\def\maketitle{
  \@author@finish
  \title@column\titleblock@produce
  \suppressfloats[t]
}
\makeatother

\begin{document}
\title{Revealing magnetism in the distorted kagome $R$Ti$_3$Bi$_4$ ($R$ = Nd, Sm, Gd) via ARPES and XMCD}

\author{C. Lim}
\email{chanyoung.lim@dipc.org}
\affiliation{Donostia International Physics Center (DIPC), Paseo Manuel de Lardizábal, 20018, San Sebastián, Spain}

\author{F. Ballester}
\affiliation{Donostia International Physics Center (DIPC), Paseo Manuel de Lardizábal, 20018, San Sebastián, Spain}
\affiliation{Department of Applied Physics, University of the Basque Country (UPV/EHU), 20018, San Sebastián, Spain}

\author{A. Kar}
\affiliation{Donostia International Physics Center (DIPC), Paseo Manuel de Lardizábal, 20018, San Sebastián, Spain}

\author{M. Alkorta}
\affiliation{Department of Applied Physics, University of the Basque Country (UPV/EHU), 20018, San Sebastián, Spain}
\affiliation{Centro de Física de Materiales (CFM) CSIC-UPV/EHU, 20018, San Sebastián, Spain}

\author{D. Subires}
\affiliation{Donostia International Physics Center (DIPC), Paseo Manuel de Lardizábal, 20018, San Sebastián, Spain}
\affiliation{University of the Basque Country (UPV/EHU), Basque Country, Bilbao, 48080 Spain}

\author{J. Dai}
\affiliation{ALBA Synchrotron Light Source, Cerdanyola del Vallès, 08290, Barcelona, Spain}

\author{M. Tallarida}
\affiliation{ALBA Synchrotron Light Source, Cerdanyola del Vallès, 08290, Barcelona, Spain}

\author{E. Vescovo}
\affiliation{National Synchrotron Light Source II, Brookhaven National Laboratory, Upton, NY, 11973, USA}

\author{T. K. Kim}
\affiliation{Diamond Light Source Ltd, Harwell Science and Innovation Campus, Didcot, OX11 0DE, UK}

\author{C. Cacho}
\affiliation{Diamond Light Source Ltd, Harwell Science and Innovation Campus, Didcot, OX11 0DE, UK}

\author{C. Yi}
\affiliation{Max Planck Institute for Chemical Physics of Solids, 01187 Dresden, Germany}

\author{S. Roychowdhury}
\affiliation{Max Planck Institute for Chemical Physics of Solids, 01187 Dresden, Germany}
\affiliation{Department of Chemistry, Indian Institute of Science Education and Research Bhopal, Bhopal-462 066, India}

\author{A. Kumar Sharma}
\affiliation{Max Planck Institute for Chemical Physics of Solids, 01187 Dresden, Germany}

\author{Y. Choi}
\affiliation{Advanced Photon Source, Argonne National Laboratory, Lemont, IL 60439, USA} 

\author{G. Fabbris}
\affiliation{Advanced Photon Source, Argonne National Laboratory, Lemont, IL 60439, USA} 

\author{J. Strempfer}
\affiliation{Advanced Photon Source, Argonne National Laboratory, Lemont, IL 60439, USA} 

\author{P. Gargiani}
\affiliation{ALBA Synchrotron Light Source, Cerdanyola del Vallès, 08290, Barcelona, Spain}

\author{C. Shekhar}
\affiliation{Max Planck Institute for Chemical Physics of Solids, 01187 Dresden, Germany}

\author{C. Felser}
\affiliation{Max Planck Institute for Chemical Physics of Solids, 01187 Dresden, Germany}

\author{I. Errea}
\affiliation{Donostia International Physics Center (DIPC), Paseo Manuel de Lardizábal, 20018, San Sebastián, Spain}
\affiliation{Department of Applied Physics, University of the Basque Country (UPV/EHU), 20018, San Sebastián, Spain}
\affiliation{Centro de Física de Materiales (CFM) CSIC-UPV/EHU, 20018, San Sebastián, Spain}

\author{M. G. Vergniory}
\affiliation{Donostia International Physics Center (DIPC), Paseo Manuel de Lardizábal, 20018, San Sebastián, Spain}
\affiliation{Département de physique et Institut quantique, Université de Sherbrooke, Sherbrooke, J1K 2R1, QC, Canada}

\author{S. Blanco-Canosa}
\email{sblanco@dipc.org}
\affiliation{Donostia International Physics Center (DIPC), Paseo Manuel de Lardizábal, 20018, San Sebastián, Spain}
\affiliation{IKERBASQUE, Basque Foundation for Science, 48013, Bilbao, Spain}

\date{\today}

\begin{abstract}
Kagome materials are known for hosting emergent quantum phenomena driven by the interaction between different lattice, charge and spin orders. Here, we present a detailed angle resolved photoemission (ARPES), density functional theory (DFT) and x-ray magnetic circular dichroism (XMCD) study of the electronic and magnetic structure of {$R$Ti$_3$Bi$_4$} ($R$ = {Nd}, {Sm}, {Gd}). ARPES and DFT demonstrate that the bulk electronic band structure is dominated by the hybridization of the Ti bands, and the weak electron-like pocket at $\Gamma$ is identified as a surface state. The isotropic XAS profile of the \textit{M}$_{4,5}$-edge of the rare earth is consistent with the presence of \textit{R}$^{3+}$ oxidation state. Using the XMCD sum rules, backed by the atomic multiplet theory calculations, we obtain the spin and orbital magnetic moments. The Ti \textit{L}$_{2,3}$-edge XMCD reveals the presence of a small magnetic moment in GdTi$_3$Bi$_4$, presumably driven by the proximity of the {Ti} kagome layers to the \textit{zigzag} chains of  Gd, while the total magnetic moment of Gd is shared by the \textit{f} and \textit{d} electrons. Our combined XMCD, ARPES and DFT study brings an important piece of information to understand the spin flip transitions and anomalous Hall effect observed in the {$R$Ti$_3$Bi$_4$} kagome metals. 

\end{abstract}

\maketitle

\renewcommand{\figurename}{Fig.}
\begin{bibunit}

\section{Introduction}
Kagome materials with the geometrical frustration arising from the corner-sharing triangles have emerged as influential platforms in the fields of condensed matter physics due to their electron correlation, topologically non-trivial bands, magnetism and intriguing interplay among them \cite{PhysRevB.45.12377,Balents2010-uz,Kang2020-rr,Chen2024-sq}. Owing to their lattice geometry, they exhibit unique features in electronic structures such as Dirac points (DPs), van Hove singularities (vHs) and flat bands \cite{PhysRevB.80.113102,Kang2020-nt}. In particular, the discoveries of AV$_3$Sb$_5$ \cite{PhysRevMaterials.3.094407} and AM$_6$X$_6$ \cite{PhysRevB.103.144410,PhysRevLett.127.266401,Wang2022-ho,Hu_undated-ed} families have had a significant impact, constituting a major focus of research in this field. Recent studies of various kagome materials have identified diverse collective phases such as charge density wave (CDW) \cite{Jiang2021-ff,PhysRevLett.129.216402,Teng2022-dm}, superconductivity \cite{PhysRevLett.125.247002,Gui2022-kb}, exotic magnetism \cite{Hou_AdvMater_2017,Ghimire_SciAdv_2020,PhysRevB.104.235139} and topological phases \cite{Ye2018-xl,Liu_Science_2019,Yin2020-qc,Li2021-hx}.

Recently, a new 134-kagome family {$R$M$_3$X$_4$} ({$R$}: rare-earth element, {M}: {V} or {Ti}, {X}: {Sb} or {Bi}) exhibits complex incommensurate magnetic order \cite{Ortiz2023-zq}, giving rise to anomalous magnetotransport properties \cite{BIE20072216,Ovchinnikov2020-sq,PhysRevMaterials.7.064201, Ortiz2023-zq,Chen2024-vz,Jiang2024-ug,Sakhya2024-wp,zheng2024anisotropic,PhysRevB.110.L121114,cheng2024striped,cheng2024spectro,bnvq-rbdc}, although their kagome lattice remains non-magnetic, unlike {AMn$_6$Sn$_6$} kagome magnets where strong magnetic interactions occur between the {Mn} kagome lattice and {A} site rare-earth elements \cite{Ghimire_SciAdv_2020, VENTURINI1996102, MALAMAN1999519, CLATTERBUCK199978}. Due to the lower symmetry, they display anisotropic kagome bands and exhibit unique magnetic topological states and rotational symmetry breakings that are not attainable in ordinary kagome materials. Therefore, the {$R$M$_3$X$_4$} compounds are potential platforms to explore unprecedented electronic structures, topological phases and to uniquely study the interplay between broken symmetry states. 

Of particular interest are the titanium-based 134-$R$Ti$_3$Bi$_4$ compounds with interleaved magnetic rare earth metals, hosting magnetic ordering \cite{Ortiz2024-yx,GUO20242660,Xinyao2025,YANG2025172916,mryp-krmp} anomalous Hall effect and rare earth induced band folding \cite{cheng2024spectro}, (in)-commensurate spin density waves (SDW) \cite{Park_2025}, which could potentially be driven by the presence of vHs bringing a high density of states (DOS) close to the Fermi level. Furthermore, the interwoven \textit{zigzag} chain accommodates magnetic rare earth metals that could magnetically polarize the itinerant TiBi kagome plane, either by orbital hybridization or the magnetically modulated Ruderman-Kittel-Kasuya-Yosida (RKKY) mechanism. This is further fueled by the observation of the strong magnetic anisotropy and stripe magnetization plateaus in the anomalous Hall effect (AHE) \cite{cheng2024striped}, presumably originated from the complex interplay between the magnetism of the rare earth and Ti sublattices. For instance, large anomalous Hall conductivity has been revealed in GdTi$_3$Bi$_4$ driven by the coexistence of skew scattering and intrinsic Berry-curvature contributions \cite{bnvq-rbdc,Singh_2026}, that intertwines with an incommensurate-commensurate charge density wave transition \cite{Han2026-xv}. Furthermore, CeTi$_3$Bi$_4$ reveals a coexistence of an incommensurate SDW and a commensurate antiferromagnetic order that emphasizes the interaction between the magnetic sublattice of Ce$^{3+}$ and the Ti kagome derived electronic structure \cite{Park_2025}. Notably, both CeTi$_3$Bi$_4$ and GdTi$_3$Bi$_4$ feature several spin flip transitions that would hint to a complex TiBi kagome and rare earth sublattice spin configuration.  

In order to disentangle the magnetic contributions responsible for the macroscopic electrodynamics reported in \textit{R}Ti$_3$Bi$_4$ kagome compounds, here we have investigated the differentiated role of the \textit{R} and Ti ions by soft x-ray absorption spectroscopy (XAS) and x-ray magnetic circular dichroism (XMCD), backed by angle resolved photoemission (ARPES), density functional theory (DFT) and atomic multiplet calculations. We explore the interplay between the rare earth metal and the Ti sublattice, given the discrepancies between the electronic structures in the (Sm,Nd,Gd)Ti$_3$Bi$_4$ kagome metals \cite{zheng2024anisotropic,PhysRevB.110.L121114,cheng2024striped,Park_2025,cheng2024spectro,Islam_2026}. We focus our study on comparing the ferromagnetic NdTi$_3$Bi$_4$ and SmTi$_3$Bi$_4$ with T$_\mathrm{C}$ = 9 K and 23 K, respectively, and the antiferromagnetic GdTi$_3$Bi$_4$, T$_\mathrm{N}$ = 13 K \cite{Ortiz2023-zq}. Through a detailed comparison between ARPES and DFT, we accurately assign the orbital character of the van Hove singularities (vHs), Dirac points (DP) and the flat bands, derived from the zero modes of the frustrated kagome plane \cite{Sakhya2024-wp} and the localized \textit{f}-electrons of the \textit{R} site responsible for the magnetism. The XAS and XMCD spectra of the rare earth are well simulated by the atomic multiplet theory, taking into account the atomic-like values of the Coulomb repulsion and the spin-orbit coupling. In addition, our XMCD reveals that the Ti sites in GdTi$_3$Bi$_4$ supports a small magnetic moment, indicating that the TiBi kagome layer is polarized by the external magnetic field at low temperatures, and the total magnetic moment of Gd is shared by the \textit{d} and more localized \textit{f} electrons. Our findings provide a piece of information to understand the electronic structure, the complex spin configuration and the anomalous Hall effect of $R$Ti$_3$Bi$_4$ kagome metals.

\section{Methods}
Single crystals of {$R$Ti$_3$Bi$_4$} ($R$={Nd}, {Sm}, {Gd}), were grown using Bi as flux, as previously reported \cite{cheng2024spectro,cheng2024striped} (see Supplementary Information S1 for a magnetic characerization of the crystals). ARPES measurements were performed at BL20 LOREA beamline of ALBA Synchrotron ({SmTi$_3$Bi$_4$}), 21-ID-1 ESM beamline of National Synchrotron Light Source II (NSLS-II), Brookhaven National Laboratory ({GdTi$_3$Bi$_4$}), and I05 beamline of Diamond Light Source ({NdTi$_3$Bi$_4$}). Single crystals of {$R$Ti$_3$Bi$_4$} were glued to the sample holder using conductive silver epoxy, with ceramic or aluminum top posts attached for the \textit{in-situ} cleaving process. ARPES measurements were carried out at the base pressure of $5 \times 10^{-11}$ torr, temperature of 20 K and under linear horizontal (LH) polarization. Total energy resolution was 25 meV or better. XAS and XMCD experiments were carried out under applied field of $\pm$6 T, at the BOREAS beamline of ALBA synchrotron ($M_{4,5}$-edge of Nd, Sm, and Gd) and at the POLAR beamline at Argonne Photon Source ($L_3$-edge of Gd). Single crystal samples were \textit{in-situ} cleaved under ultra-high vacuum condition before the measurements. The circularly polarized x-rays were incident normal and grazing (20º) to the sample surface and the measurements were taken at 2 K and ultrahigh vacuum ($7.5 \times 10^{-11}$ torr), with the magnetic field parallel to the x-ray beam. The isotropic XAS spectrum is defined as $\mu_0$=[($\mu^+$)+($\mu^-$)]/2 and the signal was recorded in the total electron yield (TEY) mode, collecting the drain current from ground to sample.  

The electronic structure of {$R$Ti$_3$Bi$_4$} was calculated by DFT using the Vienna Ab initio Simulation Package (VASP) \cite{VASP-1, VASP-2, VASP-3}. For the three compounds a symmetry preserving variable cell relaxation was performed using the structures provided in Ref. \cite{Ortiz2023-zq} as initial configurations. An energy cut-off of 400 eV was used in the calculations, together with a smearing based on the tetrahedron method with Blöch corrections, and an 8$\times$8$\times$8 gamma-centered \textit{k}-mesh for the sampling of the Brillouin zone. Generalized gradient approximation by Perdew, Burke and Ernzerhof was used for the exchange and correlation terms \cite{GGA} together with the standard projector-augmented-wave pseudopotentials included in the VASP package with the following valence configurations:  5$p^6$ 5$d^1$ 6$s^2$ for {Gd}, 5$s^2$ 5$p^6$ 5$d^1$ 6$s^2$ for {Nd}, 5$s^2$ 5$p^6$ 5$d^1$ 6$s^2$ for {Sm}, 3$p^6$ 3$d^3$ 4$s^1$ for {Ti} and 6$s^2$ 6$p^3$ for {Bi}. For the simulations, we neglected the contribution of the spin-orbit coupling term. To explore the contribution of surface states to the ARPES measurements, electronic bands were computed in a slab configuration delimited by {Ti} kagome {Bi} hexagonal surfaces. XAS and XMCD simulations were carried out using the QUANTY package \cite{Hav12,Lu14} within the local many-body atomic multiplet theory. The x-ray absorption transition at M$_{4,5}$-edge of the rare earth consists on a 3\textit{d}$\rightarrow$4\textit{f} dipole allowed transition, and the calculations take into account the spin-orbit coupling, the 3\textit{d}-4\textit{f} and 4\textit{f}-4\textit{f} Coulomb and exchange interactions, respectively, the crystal/ligand field splitting
and the hybridization with ligands.

\section{Results}
\begin{figure}
    \centering
    \includegraphics{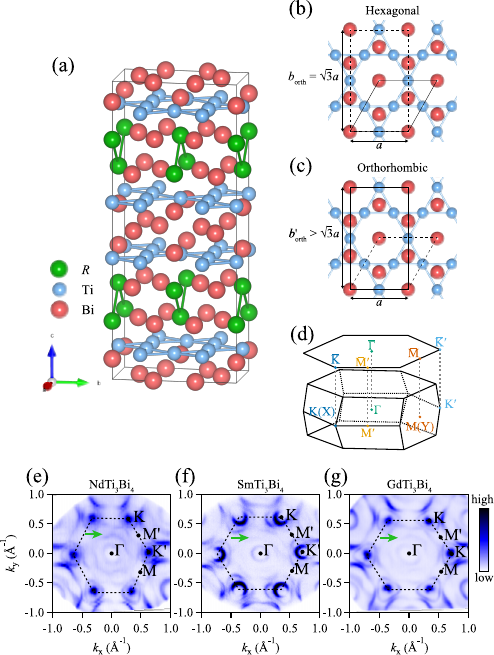}
    \caption{(a) The unit cell of {$R$Ti$_3$Bi$_4$}. The green, light blue and red spheres represent $R$, Ti and Bi atoms, respectively. (b),(c) In-plane lattice structures of conventional hexagonal kagome lattice (b) and orthorhombic kagome lattice of {$R$Ti$_{3}$Bi$_{4}$}. The solid lines in (b) and (c) show the in-plane unit cells of the hexagonal and orthorhombic lattices, respectively, while the dashed lines show the the other unit cell for reference. (d) Bulk and surface Brillouin zones of {$R$Ti$_{3}$Bi$_{4}$} compounds with the high symmetry points. (e)-(g) Fermi surfaces of (e) {NdTi$_3$Bi$_4$}, (f) {SmTi$_3$Bi$_4$}, and (g) {GdTi$_3$Bi$_4$}, measured with ARPES. The dashed lines represent the first Brillouin zones (BZs), which are nearly hexagonal. Additional Fermi surfaces marked with green arrows are observed in the $\Gamma$-{K} direction close to the $\Gamma$ pocket.}
    \label{Fig_1}
\end{figure}

The crystal structure of the {$R$Ti$_3$Bi$_4$} family is described in Fig. \ref{Fig_1}. These compounds feature in-plane \textit{zigzag} chains of $R$ atoms, hexagonal and triangular lattices of Bi atoms and slightly distorted kagome lattices of Ti atoms, resulting in an orthorhombic structure with, space group: {$Fmmm$}, No. 69. There are four Ti kagome layers in one unit cell, and the two middle layers are shifted by half a unit cell from the top and bottom layers. Between these kagome layers, {Bi} trigonal layers, {Bi} hexagonal layers, and 1D \textit{zigzag} chains of the $R$ atoms are located. Unlike other kagome compounds such as {AV$_3$Sb$_5$} or {AM$_6$X$_6$}, Fig. \ref{Fig_1}(b), the in-plane unit cell of {$R$Ti$_3$Bi$_4$}, Fig. \ref{Fig_1}(c) is orthorhombic, being the distance to the next-nearest neighbor, $b'_{orth}$, larger than $\sqrt{3}a$ due to the distorted kagome net that reduces the lattice symmetry from \textit{C}$_6$ to \textit{C}$_2$ \cite{Ortiz2023-zq}.

\begin{figure*}
    \centering
    \includegraphics{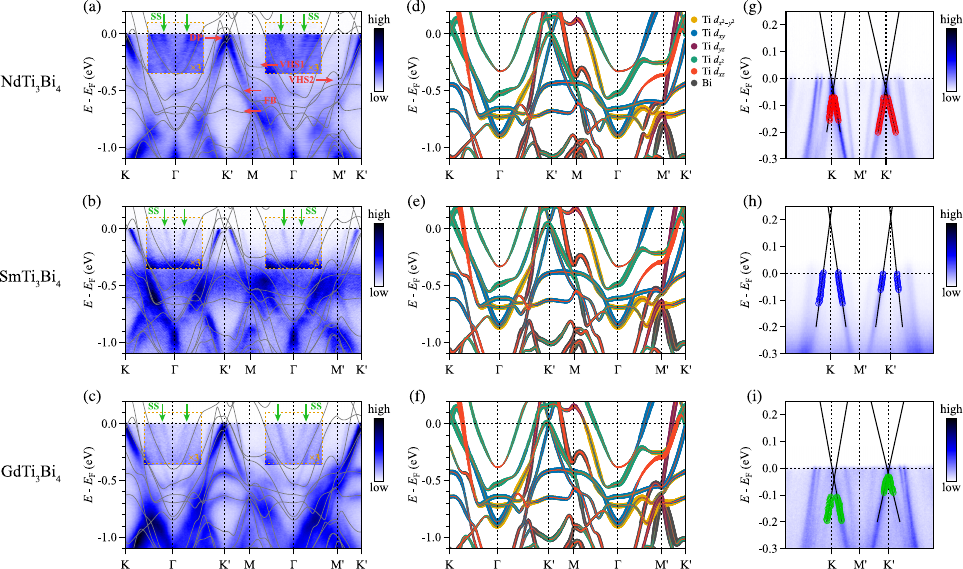}
    \caption{(a)-(c) Valence band dispersions of (a) {NdTi$_3$Bi$_4$}, (b) {SmTi$_3$Bi$_4$}, and (c) {GdTi$_3$Bi$_4$} along the high symmetry points in Figs. \ref{Fig_1}(e)-(g). Red arrows in (a) identify the electronic features characteristic of the kagome lattice; Dirac point (DP) in the vicinity of the Fermi level ({E$_\mathrm{F}$}), multiple van Hove singularities (vHs) and two flat bands. The dispersionless ARPES intensity between binding energy 0.4 and 0.5 eV in (b) is due to {Sm} {4$f$} electrons. Gray solid lines overlaid on ARPES data are dispersions from bulk DFT calculations. Green arrows mark the surface states near the $\Gamma$ points that are absent in bulk DFT calculations. Yellow dashed box regions are multiplied by 3 to better visualize the surface states. (d)-(f) DFT bulk calculation results correspond to (a)-(c) with the relative contribution from different Ti $d$ orbitals and Bi. (g)-(i) Analysis of Dirac cones at the K and K$'$ points. Momentum distribution curves (MDCs) of Dirac cones were fitted with two Lorentzian functions. Red, blue, and green markers in each plot indicate Lorentzian peak positions, and black solid lines are linear fits of the peak positions.
    }
    \label{Fig_2}
\end{figure*}

The Fermi surfaces in Figs. \ref{Fig_1}(e)-\ref{Fig_1}(g) exhibit nearly hexagonal first Brillouin zones (BZs) drawn in black dashed lines. Although the Ti kagome lattices are distorted, allowing for the distinction between {M} and {M$'$}, and {K} and {K$'$} points, the degree of elongation is less than 2$\%$ \cite{Ortiz2023-zq}, and does not influence the size of the pockets at {K} and {K$'$}. Nevertheless, while in NdTi$_3$Bi$_4$ and GdTi$_3$Bi$_4$ the DCs are located at the Fermi level, the localized \textit{f}-electrons in Sm lift the DC $\sim$0.2 eV above E$_\mathrm{F}$, see Table \ref{Tab_1}. For all three compounds, unidirectional bands near the $\Gamma$ point are observed along the $\Gamma$-{K} direction, as marked by green arrows. These features are also reproduced in the DFT-calculated Fermi surfaces (Fig. S6) and are therefore used to determine the {K} and {K$'$} points.

Figs. \ref{Fig_2}(a)-\ref{Fig_2}(c) display the overall electronic band structure of the three compounds along the high-symmetry directions. The photon energies for the $\Gamma$ and A points are determined from the $k_z$ dispersions in Fig. S3. First, we note that the valence bands (VB) present similarities, regardless of the rare-earth element $R$. The reduced symmetry of the kagome plane forces a clear anisotropy of the vHs, as the upper vHs1 band at the {M$'$} point is $\sim$0.1 eV lifted up with respect the vHs2 at M, Fig. \ref{Fig_2}(a). The kagome derived flat bands in SmTi$_3$Bi$_4$ overlap with the localized \textit{f}-electrons of Sm, Fig. \ref{Fig_2}(b), which translates into a high density of states (DOS) $\sim$0.5 eV below the Fermi level (E$_\mathrm{F}$), and hinder the accurate location of the vHs. Gd \textit{f}-electrons are located $\sim$9 eV below E$_\mathrm{F}$ (Fig. S2), while no clearly resolved Nd-derived 4$f$ spectral feature is observed. This is consistent with the expected $4f^3 \rightarrow 4f^2$ multiplet structure for Nd$^{3+}$, which leads to weak and broadly distributed spectral weight in photoemission \cite{JKLang_1981}. The anisotropy of the kagome lattice also manifests in the position of the DCs, summarized in Table \ref{Tab_1}. We note that the DC at the {K$'$} point is located at a lower binding energy than the K point, with the energy difference proportional to the degree of structural distortion from the hexagonal lattice. The DC in {NdTi$_3$Bi$_4$} and {GdTi$_3$Bi$_4$} are located near E$_\mathrm{F}$, Fig. \ref{Fig_2}(g) and (i), whereas in {SmTi$_3$Bi$_4$} are shifted approximately 0.2 eV above E$_\mathrm{F}$, as a consequence of the localized Sm \textit{f}-electrons, Fig. \ref{Fig_2}(h). Furthermore, the anisotropy of the Fermi surface is also reflected in the anisotropy of the Fermi velocity, \textit{v}$_F$ between {K} and {K$'$}, which compares the \textit{v}$_F$ to other kagome materials, Table \ref{Tab_1}, \cite{Kang2022-ws, Hu2024-gs, PhysRevB.109.035124, Lee2024-ng, Subires2025-jf, Wu2023-so}. 

\begin{table*}
\caption{Location of DPs ($E_{DP} - E_F$), Fermi velocity ($v_F$), carrier type, Dirac cone carrier density ($n$), and location of vHSs ($E_{VHs}-E_F$) of {$R$Ti$_3$Bi$_4$} compounds and other kagome compounds.}
\begin{ruledtabular}
\begin{tabular}{ccccc}
    {} &{$E_{DP}-E_F$ (eV)} &{$v_F$ ($10^5$ m/s)} &{carrier type, $n$ (cm$^{-3}$)} &{$E_{VHs}-E_F$ (eV)}\\
    
    \hline
    
    {NdTi$_3$Bi$_4$}\footnotemark[1] &{-0.044 ({K}), -0.037 ({K$'$})} &{2.88 ({K}), 2.35 ({K$'$})} &{electron, $1.218 \times 10^{19}$} &{-0.396 ({M}), -0.245 ({M$'$})}\\

    {SmTi$_3$Bi$_4$}\footnotemark[1] &{+0.197 ({K}), +0.204 ({K$'$})} &{3.32 ({K}), 3.80 ({K$'$})} &{hole, $3.392 \times 10^{19}$} &{-}\footnotemark[5]\\

    {GdTi$_3$Bi$_4$}\footnotemark[1] &{-0.040 ({K}), -0.016 ({K$'$})} &{2.62 ({K}), 2.03 ({K$'$})} &{electron, $5.376 \times 10^{18}$} &{-0.404 ({M}), -0.256 ({M$'$})}\\

    {CsV$_3$Sb$_5$}\footnotemark[2] &{-0.27, -1.08} &{3.54, 5.13} &{} &{-0.15}\\
    
    {ScV$_6$Sn$_6$}\footnotemark[3] &{-0.09, -0.28} &{2.45} &{} &{-0.02, -0.03, -0.40}\\
    
    {FeGe}\footnotemark[4] &{-0.05, -0.15, -0.65} &{2.7} &{} &{-0.26, +0.07}\\

\end{tabular}
\end{ruledtabular}

\footnotetext[1]{This work}
\footnotetext[2]{\cite{Kang2022-ws}}
\footnotetext[3]{\cite{Hu2024-gs, PhysRevB.109.035124, Lee2024-ng}}
\footnotetext[4]{\cite{Subires2025-jf, Wu2023-so}}
\footnotetext[5]{vHs are located close to the Sm 4$f$ state}

\label{Tab_1}
\end{table*}

Our DFT calculations, without considering spin-orbit coupling, nicely match the experimental band structure measured by ARPES, Figs. \ref{Fig_2}(d)-\ref{Fig_2}(f). The DFT simulations reproduce features such as flat bands resulting from the destructive interference of the electronic wavefunction, $\sim$0.5 eV below E$_\mathrm{F}$, more clearly visible in GdTi$_3$Bi$_4$, the Dirac cones near E$_\mathrm{F}$ and the multiple vHs. For all the three compounds, most of the bands near E$_\mathrm{F}$ originate from Ti 3$d$ orbitals, while the bulk Bi bands appear 1 eV below E$_\mathrm{F}$ (Fig. S5). Along the $\Gamma$-K ({K$'$}) direction, we find that the dispersive valence bands have a strong contribution from the $d_{z^2}$ orbital that hybridizes with the $d_{xy}$ orbital above $E_F$ at the {M} point, while the electron pocket at the {M} point results from $d_{xz}$ orbital character hybridized with Bi band. The DFT calculations also reveal that the $d_{xy}$-derived flat band at $\sim$-0.5 eV along the {K$'$}-{M} path hybridizes with a highly dispersive $d_{z^2}$ along {M}-{$\Gamma$} direction.

\begin{figure*}
    \centering
    \includegraphics{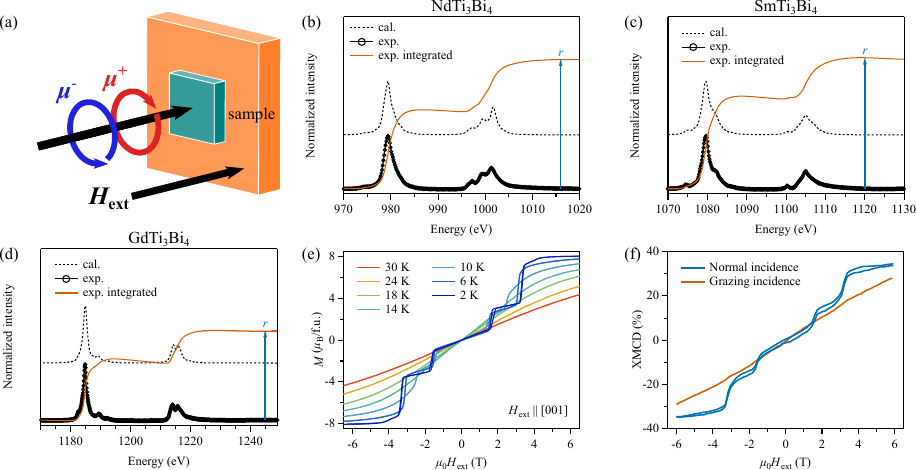}    \caption{(a) Schematic illustration of the XMCD measurement in normal geometry, with the magnetic field parallel to the beam direction. (b)-(d) Experimental (black solid line and markers) and simulated (black dashed line) isotropic XAS spectra of (b) {NdTi$_3$Bi$_4$}, (c) {SmTi$_3$Bi$_4$}, and (d) {GdTi$_3$Bi$_4$}, respectively. Red solid lines are integrated XAS of the experimental spectra. Blue arrows mark the $r$ value used for the sum rule analysis. (e),(f) Magnetic hysteresis curves of GdTi$_3$Bi$_4$ measured with (e) SQUID, and (f) XMCD. In the XMCD hysteresis curves, blue and red solid lines indicate normal and grazing incidences, respectively.}
    \label{Fig_3}
\end{figure*}

Although bulk DFT calculations reproduce most of the bands observed in ARPES measurements, they fail to capture the twofold symmetric elliptical Fermi surfaces at the $\Gamma$ point, marked with green arrows in the Fermi surface plots of Figs. \ref{Fig_1}(e)-\ref{Fig_1}(g) and the valence band plots of Figs. \ref{Fig_2}(a)-\ref{Fig_2}(c), which are also missing in the DFT-calculated bulk Fermi surfaces (Fig. S6). This variety arises from the surface bands derived from Ti and Bi terminations (Fig. S8). Considering the surface calculations, our DFT reproduces with accuracy the experimental ARPES spectra. The surface states observed here are located at the $\Gamma$ point and weakly hybridize with the bulk bands, in contrast to {YbTi$_3$Bi$_4$} and {EuTi$_3$Bi$_4$}, where the surface states are located along the $\Gamma$-{M} and $\Gamma$-{K} directions, exhibiting complex hybridization with bulk bands \cite{Sakhya2024-wp,Jiang2024-ug}. High resolution ARPES measurements revealed that this surface state electron pocket at $\Gamma$ indeed undergoes an energy splitting below the ordering temperature, directly linking the magnetism to the band splitting of the surface states \cite{zheng2024anisotropic,PhysRevB.110.L121114,c3tg-1lxl,8wmy-45m7}. Although we cannot assign the surface states to either purely Ti or Bi surface terminations, Fig. S8, we find a significant discrepancy in the {Ti} surface bands among all the three compounds, while {Bi} surface bands remained nearly identical. This behavior may be related to the crystal structure of {$R$Ti$_3$Bi$_4$}, in which the {Ti} kagome layers are located closer to the {$R$} \textit{zigzag} chains than to the {Bi} hexagonal layers in real space. This structural arrangement could suggest a possible influence of the {$R$} elements on the electronic and magnetic properties of the kagome layers.

 \begin{figure*}
    \centering
    \includegraphics{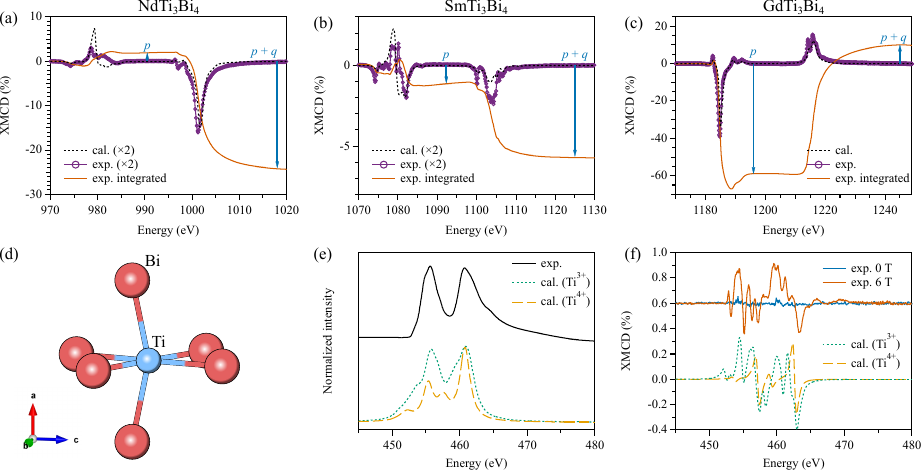}
    \caption{XMCD at the rare earth edge for (a) NdTi$_3$Bi$_4$, (b) SmTi$_3$Bi$_4$ and (c) GdTi$_3$Bi$_4$, respectively. The purple line and markers are XMCD spectra from the experiment, and the black dashed lines are spectra from the calculation. The red solid lines stand for the XMCD integration. The blue arrows mark the $p$ and $q$ values used for the sum rule analysis. (d) Coordination environment of Ti in the TiBi kagome layer, featuring a distorted polyhedra with \textit{D}$_3$\textit{h} symmetry. (e) Experimental isotropic XAS spectrum of Ti in {GdTi$_3$Bi$_4$}. The dashed spectra stand for the calculated Ti$^{3+}$ and Ti$^{4+}$ in the $D_3h$ point group. (f) XMCD spectra under 0 ad 6 T external field at the Ti edge in GdTi$_3$Bi$_4$. The dashed lines display the theoretical XMCD of Ti$^{3+}$ and Ti$^{4+}$.}
    \label{Fig_4}
\end{figure*}

To experimentally understand the intrinsic magnetic order, the interplay between rare earth metal and the TiBi sublattices, and determine its relationship with the electronic structure, we conducted XMCD experiments at the \textit{M} and \textit{L} edges of the rare earth and Ti \textit{L}-edge. In Figs. \ref{Fig_3}(b)-\ref{Fig_3}(d), we show the experimental \textit{M}$_{4,5}$ edge isotropic XAS spectra of the Nd, Sm and Gd, measured at 2 K in normal incidence geometry. The XAS of the Gd \textit{M}$_{4,5}$-edge (3\textit{d}-4\textit{f} transition), Fig. \ref{Fig_3}(d), consists on a spin-orbit split \textit{M}$_5$ and \textit{M}$_4$ levels separated by $\sim$30 eV, consistent with a 4\textit{f}$^7$ Gd$^{3+}$ oxidation state. An additional shoulder appears 2 eV below the main edge ($\sim$1185 eV) and two weak transitions around $\sim$1190 eV. The \textit{M}$_4$ edge is composed on two peaks of nearly equal intensity at a photon energies of 1215 and 1217 eV, followed by a shoulder at higher energies. Similarly, the \textit{M}$_5$ and \textit{M}$_4$ absorption edges of Sm (located at 1080 eV and 1105 eV, respectively) show weak satellites above and below the main edges. The isotropic spectrum is consistent with Sm$^{3+}$ (4\textit{f}$^5$) oxidation state, \textit{J} = $\frac{5}{2}$, without any signature of Sm$^{2+}$. Finally, the Nd \textit{M}$_{5,4}$ XAS displays the typical features of Nd$^{3+}$ 4\textit{f}$^{3}$ state, with a main \textit{M}$_5$-edge at 980 eV and 3 absorption lines at \textit{M}$_4$ ($\sim$ 1000 eV), \textit{J} = $\frac{9}{2}$.

Overlaid with the experimental XAS in Figs. \ref{Fig_3}(b)-\ref{Fig_3}(d), we also show the XAS simulations for the three compounds. Since the atomic-like character of the 4\textit{f} orbitals of the rare earths are less influenced by its local environment, we have applied a small reduction of the Slater integrals (80\% of the atomic values). The spin-orbit coupling was taken as 80\% of the atomic rare earth, and applied 1 eV Lorentzian intrinsic lifetime broadening to the \textit{M}$_{4,5}$ transitions and a Gaussian instrumental broadening of 0.1 eV. In addition, the crystal field splitting of the rare earth does not affect the XAS/XMCD calculations. As plotted in the Figs. \ref{Fig_3}(b)-\ref{Fig_3}(d), the energy profiles of Gd, Sm and Nd nicely agree with the experimental XAS, confirming their \textit{R}$^{3+}$ oxidation state and the strong localization of the \textit{f} electrons. Similarly, the XMCD spectra of the rare earth is well reproduced by the atomic multiplet calculations, see Figs. \ref{Fig_4}(a)-\ref{Fig_4}(c), after scaling by a factor of 4 to match the experimental magnetic dichroism of GdTi$_3$Bi$_4$. Moreover, the XMCD hysteresis loop carried out at the \textit{M}$_5$-edge of Gd, \ref{Fig_3}(f), nicely follows the spin flip transitions observed in the magnetization measurements, Fig. \ref{Fig_3}(e). We detect a small out of plane magnetic dichroic signal at the Ti \textit{L}-edge in GdTi$_3$Bi$_4$, see Fig. \ref{Fig_4}(f). The XMCD is fully suppressed at 0 T and consistent with the absence of dichroism of antiferromagnetic Gd at 0 T, indicating that the spin polarization of Ti follows the magnetization of Gd under magnetic field. 

The experimental magnetic moment of the rare earths were obtained from the well-known sum rules, performed upon both XAS and XMCD spectra. The orbital and spin magnetic moments were quantified by using the following equations\cite{PhysRevLett.68.1943,PhysRevLett.70.694}:

\begin{equation}
    \label{eq1}
    \begin{split}
          M_{orb}^{R} &=-2\frac{n_{h}^{R}}{r}\int_{M_4+M_5}(\mu^+-\mu^-)\ dE\\
          \\
          &=-\frac{2n_{h}^{R}}{r}(p+q)
    \end{split}
\end{equation}

and

\begin{multline}
        \label{eq2}
        M_{spin,eff}^{R} = 2<S_z> +  6<T_z> \\= -\frac{n_{h}^{R}}{r}\left[2\int_{M_5}(\mu^+-\mu^-)\ dE
         -3\int_{M_4}(\mu^+-\mu^-)\ dE\right]\\=-\frac{2n_{h}^{R}}{r}\ (p-\frac{3}{2}q)
\end{multline}

where \textit{n}$_{h}^{R}$ is the number of holes in the \textit{f} state (\textit{n}$_{h}^{Sm}$=9, \textit{n}$_{h}^{Nd}$=11, \textit{n}$_{h}^{Gd}$=7). The \textit{r} parameter obtained from the isotropic XAS spectrum, previously corrected by a step function, and the values of \textit{p} and \textit{q} are defined in Figs. \ref{Fig_4}(a)-\ref{Fig_4}(c). For Sm, the off-diagonal term of the 3\textit{d}–4\textit{f} exchange interaction causes a large \textit{jj} mixing, restricting the applicability of the sum rules. The orbital and spin magnetic moments of Nd are 0.55$\pm$0.01$\mu_\mathrm{B}$ and 1.87$\pm$0.01$\mu_\mathrm{B}$, respectively, very close to the total magnetic moment reported in the literature \cite{Ortiz2023-zq}. On the other hand, the magnetic moment of the \textit{f} electrons of Gd amount to $\sim$7.03$\mu_\mathrm{B}$ at 6T (0.21$\pm$0.02$\mu_\mathrm{B}$ of orbital and 6.82$\pm$0.06$\mu_\mathrm{B}$ of spin moments), lower than the ordered magnetic moment measured in magnetization, $\sim$7.9$\mu_\mathrm{B}$, see Fig. \ref{Fig_3}(e). Sweeping the XMCD through the Gd \textit{L}$_3$-edge, that corresponds to the 2\textit{p}-5\textit{d} transition, see Fig. S9, we find $\sim$0.42 $\mu_\mathrm{B}$ hosted by the 5\textit{d} electrons, lending $\sim$0.4$\mu_\mathrm{B}$ to the TiBi sublattice, provided that the mixing of the \textit{L}$_3$ and \textit{L}$_2$ edges prevents us from the direct application of the sum rules at the Ti \textit{L}-edge. Finally, no dichroism is detected at the Ti-edge in NdTi$_3$Bi$_4$ and SmTi$_3$Bi$_4$, despite the localized Sm bands strongly overlap with the Ti bands $\sim$0.3 eV below E$_\mathrm{F}$.   

\section{Discussion}

The \textit{R}Ti$_3$Bi$_4$ compounds exhibit similar electronic structures, including hexagonal Fermi surfaces, DCs, vHs, and flat bands. In comparison with previous reports, our combined ARPES measurements and DFT calculations allow us to clarify the orbital character of these bands and thus provide a more complete description of the electronic structure. Furthermore, we present a quantitative analysis of the anisotropy of the Dirac cones at the {K} and {K$'$} points. We also identify electron-like surface states at the $\Gamma$ point, derived from both Ti and Bi layers, further enriching the overall picture. In this context, we note the recent STM study reporting an unconventional spin-intertwined CDW in {GdTi$_3$Bi$_4$} \cite{Han2026-xv}. The $\Gamma$-centered surface states identified here may provide a natural electronic basis for a CDW instability. Their presence at low energies suggests that they could actively participate in the electronic reconstruction observed by STM.

Our XMCD results indicate that the \textit{f} electrons of the rare earth dominate the magnetism in \textit{R}Ti$_3$Bi$_4$, presenting a spin and orbital magnetic moment very close to the measured bulk values, and nicely follow the macroscopic magnetic behavior at low temperatures, see Fig. \ref{Fig_3}(e) and \ref{Fig_3}(f). Nevertheless, the total magnetic moment (spin and orbital) derived from the highly localized Gd 4\textit{f} states of Gd is lower than the total magnetization measured experimentally in GdTi$_3$Bi$_4$. Our comprehensive XMCD data solves this discrepancy by assigning $\sim$0.4 $\mu_\mathrm{B}$ to the \textit{d} electrons of Gd and $\sim$0.1$\mu_\mathrm{B}$/Ti at the Ti sites, presumably as a result of the small spin polarization along the applied field direction that couples to the out-of-plane magnetic moments of Gd. Although, surface magnetism has been widely discussed in the context of 2D electron gas \cite{Lee_2013,Salluzzo_2013}, intermetallic compounds \cite{PhysRevLett.94.146403,PhysRevLett.98.037202}, {$R$Rh$_2$Si$_2$} antiferromagnets \cite{Chikina_2014,Guttler2016-gs} and quantum electron liquids \cite{Kim_2022}, this mechanism is unlikely, since Ti XMCD is absent in SmTi$_3$Bi$_4$ and NdTi$_3$Bi$_4$ within our detection limit. This leads us to assume that the proximity of the {Ti} kagome layers to the $R$ \textit{zigzag} chains spin polarize the Ti bands. Indeed, the strong polarization of the \textit{d} states by the presence of the localized \textit{f} magnetic moments at the Gd atoms via the dipolar 4\textit{f}-5\textit{d} exchange interaction can, in turn, polarize the electronic states of Ti via Gd 5\textit{d}-Ti 3\textit{d} hybridization \cite{Sarafidis_2009}. To confirm this hypothesis, we present the simulation of the XMCD spectrum of Ti, assuming a distorted octahedral chemical environment of the Bi-sites of Ti, Fig. \ref{Fig_4}(d). For the XMCD calculations, we approximated the local symmetry of Ti to the \textit{D}$_3$\textit{h} point group and assume the atomic value of Ti for the spin-orbit coupling, $\zeta_{2p}$=3.776 eV and the crystal field parameters $\Delta_\mu$ and $\Delta_\nu$ of 0.3 and -0.3 eV, chosen to match the experimental XMCD spectrum, see SI Fig. S10. Notably, the experimental XMCD is consistent with the spectrum containing Ti$^{3+}$ oxidation state, instead of purely Ti$^{4+}$, as expected from the electron counting approach, presumably due to a weak charge leak from Bi to Ti. Although the simulated spectra had to be scaled, the XMCD qualitatively agrees with the experimental one, in particular the double peak structure at the \textit{L}$_3$ and \textit{L}$_2$ edges, highlighted as shadowed regions in Fig. \ref{Fig_4}(f), that is not captured by XMCD spectrum of Ti$^{4+}$.

Finally, we would like to connect these findings to the physics of selected {\textit{R}Ti$_3$Bi$_4$} compounds. The small magnetic moment of Ti aligns collinearly with the ordered magnetic moments of Gd that  contribute to the observed anomalous Hall effect in GdTi$_3$Bi$_4$ and EuTi$_3$Bi$_4$ \cite{cheng2024striped,PhysRevB.111.155103,bnvq-rbdc}. One possibility is that the strong magnetic moment of the localized 4\textit{f} electrons of such rare-earth elements interact with non-magnetic kagome planes via direct- or indirect-exchange interaction or RKKY interaction, breaking time-reversal symmetry or opening gaps in exotic surface states, and leading to quantum anomalous Hall effect \cite{Wei2016GrapheneEuS,Katmis2016TopologicalEuS,Jiang2015EuSBi2Se3,RKKYClassic,Yang2013proximitygraphene}.

\section{Conclusion}
In conclusion, we have conducted a detailed study of the electronic structure and magnetism of {$R$Ti$_3$Bi$_4$} ($R$ = {Nd}, {Sm}, {Gd}), using ARPES, XMCD and DFT. We have achieved a comprehensive description of electronic band structure, both at the ARPES and DFT level, and capture the orbital character of the bands and the surface states of the energy spectrum. Moreover, we reveal a small magnetic moment in the Ti atomic sites in GdTi$_3$Bi$_4$. The out-of-plane dichroism at the Ti \textit{L}-edge is presumably driven by the proximity of the {Ti} kagome layers to the Gd \textit{zigzag} chains that polarize the Ti surface bands.

\section*{Acknowledgments}
The ARPES experiments in this work were conducted at the BL 20 LOREA beamline at ALBA Synchrotron, 21-ID ESM of the National Synchrotron Light Source II, a U.S. Department of Energy (DOE) Office of Science User Facility operated for the DOE Office of Science by Brookhaven National Laboratory under Contract No. DE-SC0012704, and instrument I05 at Diamond Light Source. LOREA beamline is co-funded by the European Regional Development Fund (ERDF) within the ”Framework of the Smart Growth Operative Programme 2014-2020. D.S., A. Kar and S.B-C. thank to the MINECO of Spain through the projects PID2021-122609NB-C21 and PID2024-161503NB-C21 and by the European Union Next Generation EU/PRTR-C17.I1, as well as by IKUR Strategy under the collaboration agreement between Ikerbasque Foundation and DIPC on behalf of the Department of Education of the Basque Government. C.L. is supported by the European Research Council (ERC) under the European Union’s Horizon 2020 research and innovation program (grant agreement no. 101020833). M.G.V received financial support from the Canada Excellence Research Chairs Program for Topological Quantum Matter.  M.G.V and F.B. thank support to the Spanish Ministerio de Ciencia e Innovacion grant PID2022-142008NB-I00 and the Ministry for Digital Transformation and of Civil Service of the Spanish Government through the QUANTUM ENIA project call - Quantum Spain project, and by the European Union through the Recovery, Transformation and Resilience Plan - NextGenerationEU within the framework of the Digital Spain 2026 Agenda. I.E. also acknowledges financial support from the Department of Education, Universities and Research of the Eusko Jaurlaritza, and the University of the Basque Country UPV/EHU (Grant No. IT1527-22); and the Spanish Ministerio de Ciencia e Innovación (Grant No. PID2022-142861NA-I00). C.Y., A.S., C.S., and C.F. acknowledge financial support by the Deutsche Forschungsgemeinschaft  (DFG, German Research Foundation) through the Würzburg-Dresden Cluster of Excellence ctd.qmat – Complexity, Topology and Dynamics in Quantum  Matter (EXC 2147, project-id 390858490). P.G. acknowledge support from the MINECO project PID2023-149494NB-C32. This work was carried out with the support of Diamond Light Source, instrument I05-ARPES (proposal SI36505), and used resources (beamline 4ID) of the Advanced Photon Source, a U.S. DOE Office of Science User Facility operated for the DOE Office of Science by Argonne National Laboratory under Contract No. DE-AC02-06CH11357. 

\putbib[references_RETi3Bi4]
\end{bibunit}

\clearpage
\onecolumngrid

\setcounter{section}{0}
\renewcommand{\thesection}{\arabic{section}}

\renewcommand\figurename{Fig. S}
\renewcommand\tablename{Table. S}
\makeatletter
\def\fnum@figure{\figurename\thefigure}
\def\fnum@table{\tablename\arabic{table}}
\makeatother

\title{Supplementary Material of `Revealing magnetism in the distorted kagome $R$Ti$_3$Bi$_4$ ($R$ = Nd, Sm, Gd) via ARPES and XMCD'}
\author{C. Lim}
\email{chanyoung.lim@dipc.org}
\affiliation{Donostia International Physics Center (DIPC), Paseo Manuel de Lardizábal, 20018, San Sebastián, Spain}

\author{F. Ballester}
\affiliation{Donostia International Physics Center (DIPC), Paseo Manuel de Lardizábal, 20018, San Sebastián, Spain}
\affiliation{Department of Applied Physics, University of the Basque Country (UPV/EHU), 20018, San Sebastián, Spain}

\author{A. Kar}
\affiliation{Donostia International Physics Center (DIPC), Paseo Manuel de Lardizábal, 20018, San Sebastián, Spain}

\author{M. Alkorta}
\affiliation{Department of Applied Physics, University of the Basque Country (UPV/EHU), 20018, San Sebastián, Spain}
\affiliation{Centro de Física de Materiales (CFM) CSIC-UPV/EHU, 20018, San Sebastián, Spain}

\author{D. Subires}
\affiliation{Donostia International Physics Center (DIPC), Paseo Manuel de Lardizábal, 20018, San Sebastián, Spain}
\affiliation{University of the Basque Country (UPV/EHU), Basque Country, Bilbao, 48080 Spain}

\author{J. Dai}
\affiliation{ALBA Synchrotron Light Source, Cerdanyola del Vallès, 08290, Barcelona, Spain}

\author{M. Tallarida}
\affiliation{ALBA Synchrotron Light Source, Cerdanyola del Vallès, 08290, Barcelona, Spain}

\author{E. Vescovo}
\affiliation{National Synchrotron Light Source II, Brookhaven National Laboratory, Upton, NY, 11973, USA}

\author{T. K. Kim}
\affiliation{Diamond Light Source Ltd, Harwell Science and Innovation Campus, Didcot, OX11 0DE, UK}

\author{C. Cacho}
\affiliation{Diamond Light Source Ltd, Harwell Science and Innovation Campus, Didcot, OX11 0DE, UK}

\author{C. Yi}
\affiliation{Max Planck Institute for Chemical Physics of Solids, 01187 Dresden, Germany}

\author{S. Roychowdhury}
\affiliation{Max Planck Institute for Chemical Physics of Solids, 01187 Dresden, Germany}
\affiliation{Department of Chemistry, Indian Institute of Science Education and Research Bhopal, Bhopal-462 066, India}

\author{A. Kumar Sharma}
\affiliation{Max Planck Institute for Chemical Physics of Solids, 01187 Dresden, Germany}

\author{Y. Choi}
\affiliation{Advanced Photon Source, Argonne National Laboratory, Lemont, IL 60439, USA} 

\author{G. Fabbris}
\affiliation{Advanced Photon Source, Argonne National Laboratory, Lemont, IL 60439, USA} 

\author{J. Strempfer}
\affiliation{Advanced Photon Source, Argonne National Laboratory, Lemont, IL 60439, USA} 

\author{P. Gargiani}
\affiliation{ALBA Synchrotron Light Source, Cerdanyola del Vallès, 08290, Barcelona, Spain}

\author{C. Shekhar}
\affiliation{Max Planck Institute for Chemical Physics of Solids, 01187 Dresden, Germany}

\author{C. Felser}
\affiliation{Max Planck Institute for Chemical Physics of Solids, 01187 Dresden, Germany}

\author{I. Errea}
\affiliation{Donostia International Physics Center (DIPC), Paseo Manuel de Lardizábal, 20018, San Sebastián, Spain}
\affiliation{Department of Applied Physics, University of the Basque Country (UPV/EHU), 20018, San Sebastián, Spain}
\affiliation{Centro de Física de Materiales (CFM) CSIC-UPV/EHU, 20018, San Sebastián, Spain}

\author{M. G. Vergniory}
\affiliation{Donostia International Physics Center (DIPC), Paseo Manuel de Lardizábal, 20018, San Sebastián, Spain}
\affiliation{Département de physique et Institut quantique, Université de Sherbrooke, Sherbrooke, J1K 2R1, QC, Canada}

\author{S. Blanco-Canosa}
\email{sblanco@dipc.org}
\affiliation{Donostia International Physics Center (DIPC), Paseo Manuel de Lardizábal, 20018, San Sebastián, Spain}
\affiliation{IKERBASQUE, Basque Foundation for Science, 48013, Bilbao, Spain}

\date{\today}
\begin{bibunit}

\maketitle

\clearpage
\onecolumngrid

\section{Magnetic characterization}

Magnetization measurements of {$R$Ti$_3$Bi$_4$} ($R$=Nd, Sm and Gd), see Fig. S1. The temperature dependence of the magnetization reveals magnetic transitions of $\sim$10 K (Nd), $\sim$25 K (Sm) and $\sim$15 K (Gd), in agreement with the reports in the literature \cite{zheng2024anisotropic,PhysRevB.110.L121114,cheng2024striped,Park_2025,cheng2024spectro,Islam_2026}.   

\begin{figure}[htbp]
    \centering
    \includegraphics{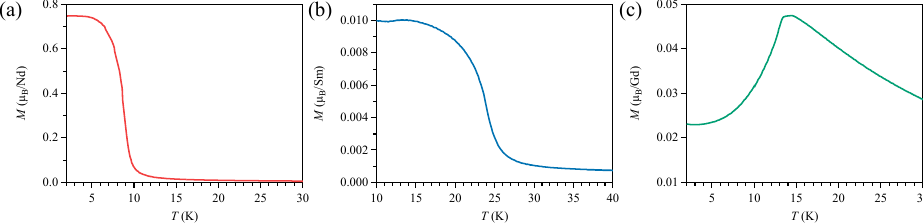}
    \caption{Temperature-dependent magnetization of (a) {NdTi$_3$Bi$_4$}, (b) {SmTi$_3$Bi$_4$}, and (c) {GdTi$_3$Bi$_4$}}
    \label{Fig_S1}
\end{figure}

\section{Core level XPS}
\begin{figure}
    \centering
    \includegraphics{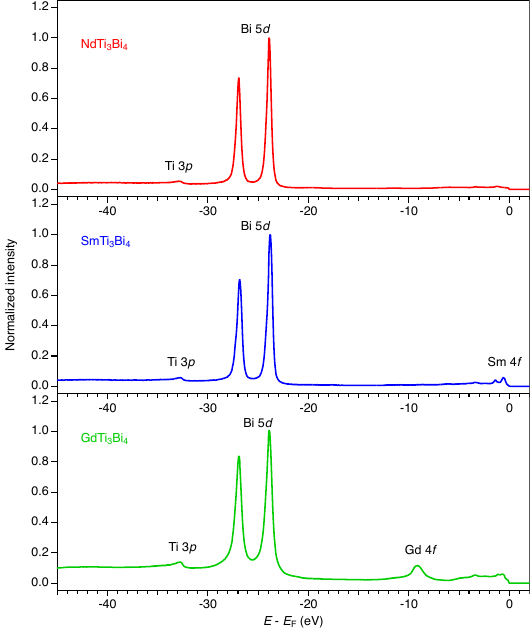}
    \caption{Core level spectra of {NdTi$_3$Bi$_4$} (top, red), {SmTi$_3$Bi$_4$} (middle, blue), and {GdTi$_3$Bi$_4$} (bottom, green). All intensities were normalized to their maximum value.}
    \label{Fig_S2}
\end{figure}

Before the ARPES measurements, the core-level spectra of {$R$Ti$_3$Bi$_4$} compounds ({$R$} = {Nd}, {Sm}, and {Gd}) were examined using 110 eV photons. Fig. S\ref{Fig_S2} shows the resulting core-level spectra, which reveal {Bi} 5$d$ and {Ti} 3$p$ peaks common to all three compounds. In contrast, each compound exhibits distinct lanthanide 4$f$ spectral features. In {SmTi$_3$Bi$_4$}, the 4$f$ peak appears near the Fermi level ({E$_F$}), consistent with the ARPES results, whereas in {GdTi$_3$Bi$_4$} it is located 9 eV below {E$_F$}. For {NdTi$_3$Bi$_4$}, no clear 4$f$-related peak is observed, due to $4f^3 \rightarrow 4f^2$ multiplet structure for Nd$^{3+}$, which leads to weak and broadly distributed spectral weight in photoemission.

\section{angle resolved photoelectron spectroscopy}

\begin{figure}[htbp]
    \centering
    \includegraphics{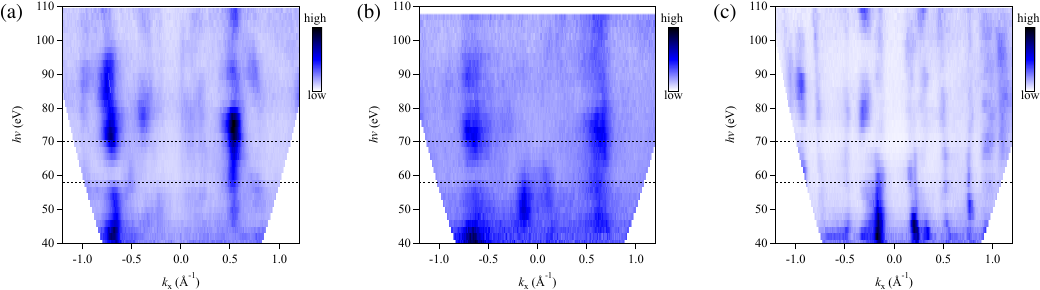}
    \caption{$k_x$-$h\nu$ plane Fermi surfaces of (a) {NdTi$_3$Bi$_4$} , (b) {SmTi$_3$Bi$_4$} , and (c) {GdTi$_3$Bi$_4$} . Dashed lines indicate the locations of $\Gamma$ (70 eV) and A (58 eV) planes.}
    \label{Fig_S3}
\end{figure}

Fig. S\ref{Fig_S3} displays $k_x$-$h\nu$ plane Fermi surfaces of {$R$Ti$_3$Bi$_4$} compounds, which correspond to the $k_x$-$k_z$ plane Fermi surfaces. The  two-dimensional nature of these Fermi surfaces confirms minimal interlayer electronic interactions in these compounds.

\begin{figure}
    \centering
    \includegraphics{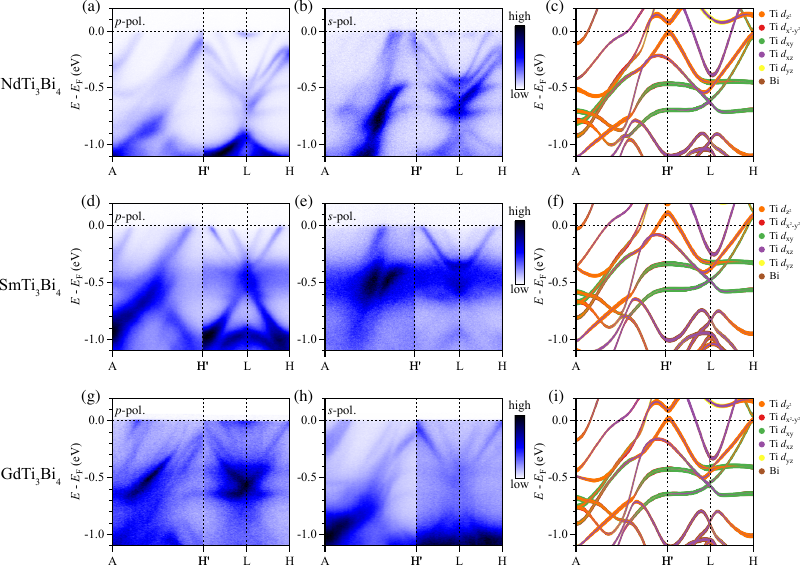}
    \caption{{A}-{H'}-{L}-{H} dispersions of {NdTi$_3$Bi$_4$} ((a)-(c)), {SmTi$_3$Bi$_4$} ((d)-(f)) and {GdTi$_3$Bi$_4$} ((g)-(i)). (a),(d), and (g) show ARPES results under $s$-polarization, while (b),(e), and (h) correspond to ARPES results under $p$-polarization. (c),(f), and (i) present DFT bulk calculation results with the relative contribution from different Ti $d$ orbitals and Bi.
    }
    \label{Fig_S4}
\end{figure}

Fig. S\ref{Fig_S4} compares the A-plane electronic structures obtained from ARPES and DFT. As shown in Fig. S\ref{Fig_S3}, these materials exhibit two-dimensional electronic structures, and the A plane electronic structure closely resemble the $\Gamma$ plane discussed in the main text. However, unlike in the $\Gamma$ plane, an electron band at the Brillouin zone center is observed in the A plane in both ARPES and DFT. Taking this into account, the electronic structures measured by ARPES are in good agreement with those calculated by DFT.

\section{dft calculations}

For the pDOS calculation we used a 21$\times$21$\times$21 $\Gamma$-centered k-mesh grid for the BZ sampling, using the same parameters as with the calculations mentioned in the main text. The valence states are dominantly Ti 3$d$, with a noticeable contribution from Bi states starting from approximately 1 eV below $E_F$. Between the three compounds, we found no significant difference between the occupation of the rare-earth $d$ orbitals.

\begin{figure}
    \centering
    \includegraphics{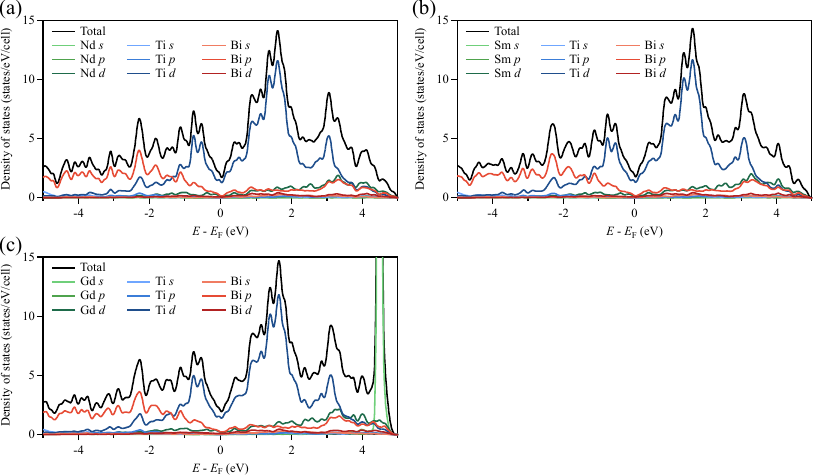}
    \caption{Calculated partial density of states (pDOS) of (a) {NdTi$_3$Bi$_4$}, (b) {SmTi$_3$Bi$_4$}, and (c) {GdTi$_3$Bi$_4$}. Each color represents a different orbital contribution, as indicated in each plot.}
    \label{Fig_S5}
\end{figure}

\begin{figure}
    \centering
    \includegraphics{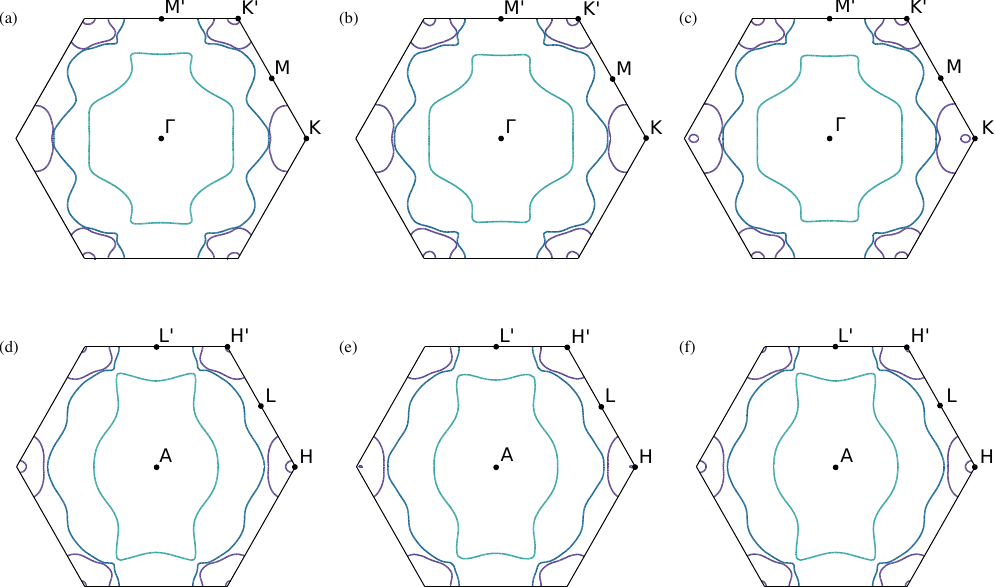}
    \caption{DFT-calculated bulk Fermi surfaces for the $\Gamma$ plane ((a)-(c)) and the A plane ((d)-(f). Panels (a), (d) correspond to {NdTi$_3$Bi$_4$}, (b), (e) to {SmTi$_3$Bi$_4$}, and (c), (f) to {GdTi$_3$Bi$_4$}, respectively. }
    \label{Fig_S6}
\end{figure}

In addition to the band dispersions, we calculated the Fermi surfaces of the $\Gamma$ plane and the A plane for the {$R$Ti$_3$Bi$_4$} compounds (Fig. S\ref{Fig_S6}). The two distinct {K} and {M'} ({H} and {L'}) points and the four {K'} and {M} ({H'} and {L}) points are well reproduced, in good agreement with the ARPES results. However, in contrast to the Fermi surfaces observed by ARPES in Fig. 1(e)-1(f), the pocket at the Brillouin-zone center is not captured in the bulk DFT calculations.

\begin{figure}
    \centering
    \includegraphics{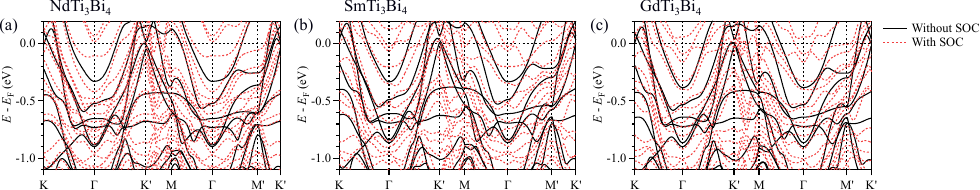}
    \caption{DFT-calculated electronic structures of (a) {NdTi$_3$Bi$_4$}, (b) {SmTi$_3$Bi$_4$}, and (c) {GdTi$_3$Bi$_4$}. Black solid lines correspond to calculations without SOC, while red dashed lines represent calculations including SOC.}
    \label{Fig_S7}
\end{figure}

The band structure calculations including SOC were computed using the same parameters as the bands shown in the main text. In Fig. S\ref{Fig_S7}, we present a comparison of the band structures of the three compounds with and without SOC included in the DFT calculations. Previous studies \cite{Islam_2026} suggest that the effect of SOC is generally weak in similar rare-earth kagome systems, except in the vicinity of Dirac points. In our case, the inclusion of SOC leads to noticeable modifications of the band dispersions over a broader energy range, which are not clearly reflected in the ARPES measurements. This may indicate that the SOC effect is overestimated within the present DFT framework for these compounds. On the other hand, prior works \cite{Island2019-ay, ZhangPRB2020} have shown that SOC can induce band gap openings and influence band topology near Dirac points and flat bands in kagome materials. Taken together, these results suggest that while SOC may play a role in shaping local band features, its overall impact on the electronic structure of {$R$Ti$_3$Bi$_4$} appears to be limited in describing the experimentally observed dispersions.

\begin{figure}
    \centering
    \includegraphics{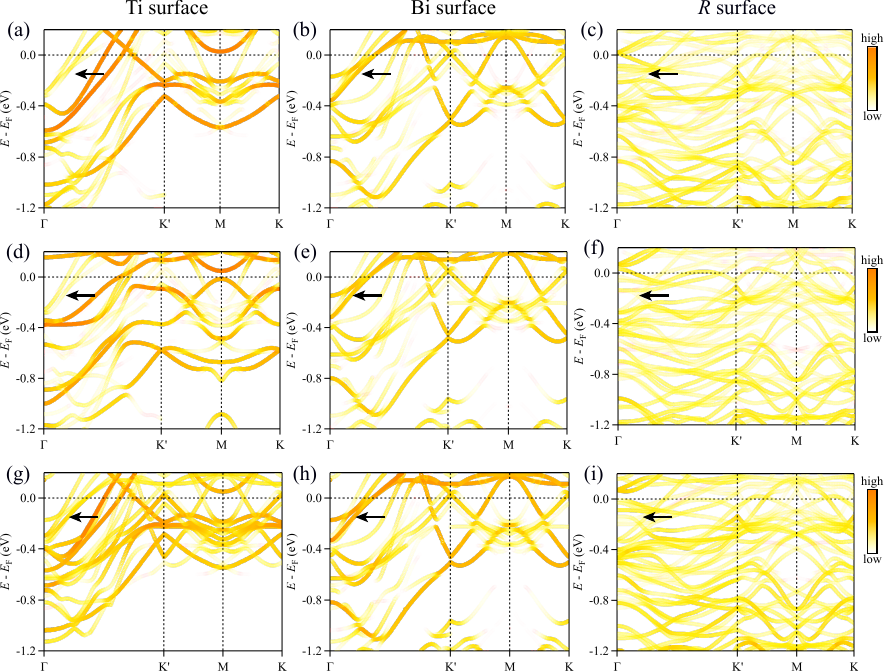}
    \caption{DFT calculation results for surface bands of {NdTi$3$Bi$4$} ((a)-(c)), {SmTi$3$Bi$4$} ((d)-(f)), and {GdTi$3$Bi$4$} ((g)-(i)). (a), (d), and (g) correspond to surface bands for Kagome Ti termination. (b), (e), and (h) correspond to surface bands for trigonal Bi termination. (c), (f), and (i) correspond to rare-earth termination. Black arrows mark the bands which are observed in ARPES and absent in bulk DFT calculations.}
    \label{Fig_S8}
\end{figure}

Based on these results, we performed surface-band DFT calculations, assuming that the observed bands at $\Gamma$ originate from surface states (Fig. S\ref{Fig_S8}). Except for the Sm 4$f$ states, the valence states of the {$R$Ti$_3$Bi$_4$} compounds are mainly composed of {Ti}- and {Bi}-derived electrons, as shown in Fig. 2, Fig. S\ref{Fig_S4} and Fig. S\ref{Fig_S5}. Accordingly, surface-band calculations were carried out for both {TiBi} kagome layer termination and {Bi} hexagonal layer termination, where these electronic states are expected to be most prominent. In addition, rare-earth terminations were also included based on preferred terminations identified in previous studies \cite{PhysRevB.110.L121114}. As a result, we identified a $\Gamma$ surface electron pocket that is absent in the bulk DFT calculations. However, since this band appears for both {Ti} and  {Bi} terminations, although it is slightly more pronounced for the Bi termination, we cannot unambiguously determine its origin. Moreover, while the {Bi}-terminated surfaces show little variation among different compounds, the {TiBi}-terminated surfaces exhibit significant changes depending on the rare-earth element {$R$}. This behavior suggests an interaction between the {$R$} \textit{zigzag} chains and the {TiBi} kagome layers. Unlike {TiBi} and {Bi} terminations, rare-earth terminations did not exhibit a distinct surface band, which appears to be related to the low pDOS in the valence state of rare-earth atoms.

\section{xmcd at the {Gd} l-edge}

\begin{figure}
    \centering
    \includegraphics{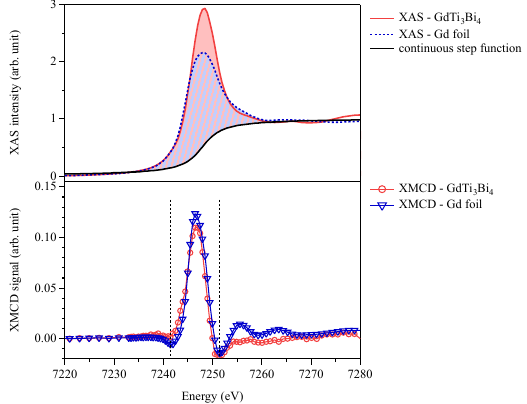}
    \caption{Gd $L_3$-edge XAS spectra of {GdTi$_3$Bi$_4$} and Gd foil (top) and the corresponding XMCD spectra (bottom). Red and blue areas are XAS areas of {GdTi$_3$Bi$_4$} and Gd foil, after removing the continuous step function drawn in black solid line. Black dashed lines indicate the energy range used for the estimation of 5$d$ magnetic moment in {GdTi$_3$Bi$_4$} (7241.4 and 7251.4 eV).
    }
    \label{Fig_S9}
\end{figure}

Fig. S\ref{Fig_S9} shows the estimation of the magnetic moment from Gd 5$d$ electrons using {Gd} $L$-edge XMCD results (2\textit{p}-5\textit{d} transition). First, we measured XAS and XMCD spectra at the {Gd} $L_3$ edge for both {Gd} foil and then calculate the magnetic moment in {GdTi$_3$Bi$_4$}. It is known that the {Gd} 5$d$ electrons responsible for the L edge typically carry a magnetic moment of 0.6 $\pm$ 0.12 $\mu_B$\cite{PhysRevLett.115.096402}. The 5$d$ electron moment of {GdTi$_3$Bi$_4$} was estimated by comparing the XMCD and XAS $L_3$ edges with pure Gd foil, since the XMCD signal is proportional to the 5$d$ magnetic moment. The foil was measured in transmission and the crystal in fluorescence mode at 10 K with 6 T. The x-ray direction (parallel to the field direction) as 10 degrees of the surface normal direction for the crystal. Here, XAS is defined as the average of the two XAS with opposite circumpolar polarizations, and XMCD as the difference. While the XAS area for the {GdTi$_3$Bi$_4$} crystal (red area in the top graph of Fig. S\ref{Fig_S9}) is about 1.2 times than that of Gd foil (blue area in the top graph of Fig. S\ref{Fig_S9}), the XMCD area between the dashed lines for {GdTi$_3$Bi$_4$} is 0.84 of the foil value. Thus, normalizing by the XAS area, the XMCD signal of {GdTi$_3$Bi$_4$} is 0.84 / 1.2 = 0.7 of the foil. Since the Gd foil has 0.6 $\mu$B/Gd in the 5$d$ orbital, then the estimated value for the {GdTi$_3$Bi$_4$} crystal is 0.7 $\times$ 0.6 $\mu$B = 0.42 $\mu$B

\section{xmcd simulation of the {Ti} l-edge}

\begin{figure}[htbp]
    \centering
    \includegraphics{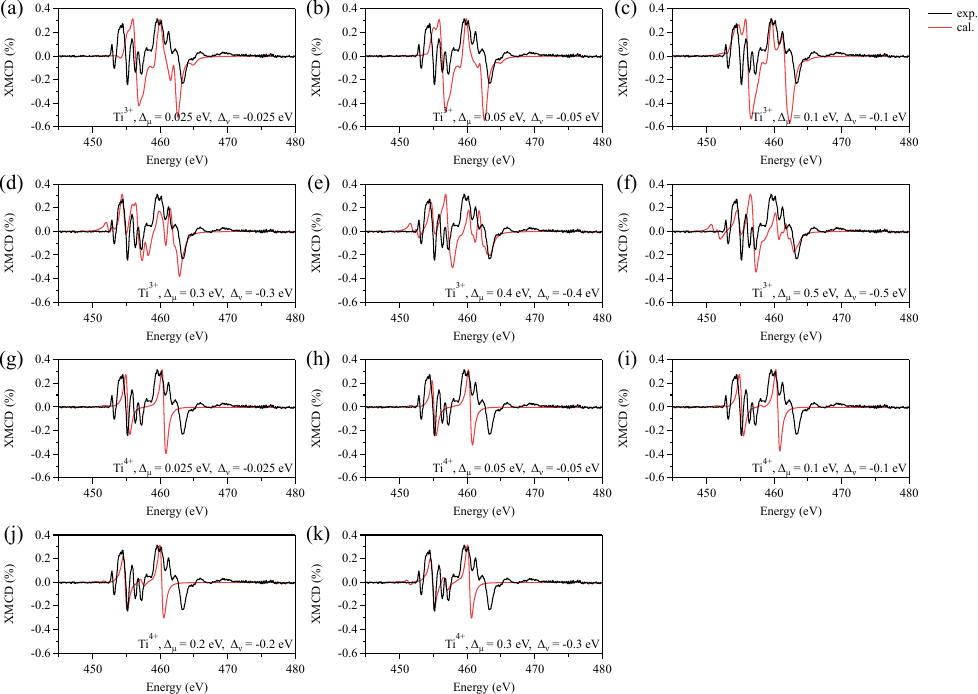}
    \caption{Simulation of the {Ti$^{3+}$} and {Ti$^{4+}$} \textit{L}-edge XMCD as a function of the $\Delta_\mu$ and $\Delta_\nu$ within the \textit{D}$_3$\textit{h} point group.}
    \label{Fig_S10}
\end{figure}

Fig. \ref{Fig_S10} compares the XMCD of the Ti$^{3+}$ and Ti$^{4+}$ within the \textit{D}$_3$\textit{h} point group for different values of the crystal field parameters $\Delta_\mu$ and $\Delta_\nu$. The experimental spectra is consistent with the presence of Ti$^{3+}$, rather than Ti$^{4+}$. Taking $\Delta_\mu$ and $\Delta_\nu$ for Ti$^{3+}$ as +0.3 and -0.3 eV, respectively, qualitatively reproduces the experimental XMCD spectra. On the other hand, the simulated Ti$^{4+}$ XMCD spectra is barely dependent on the crystal field parameters.

\putbib[references_RETi3Bi4]
\end{bibunit}
\end{document}